\documentclass[conference]{IEEEtran}
\usepackage[utf8]{inputenc}
\IEEEoverridecommandlockouts
\usepackage{cite}
\usepackage{amsmath,amssymb,amsfonts}
\usepackage{algorithmic}
\usepackage{graphicx}

\usepackage{multirow}
\usepackage[table,xcdraw]{xcolor}

\usepackage{textcomp}
\usepackage{comment}
\usepackage{xcolor}
\usepackage{amsmath,amssymb,amsfonts}
\usepackage{algorithmic}
\usepackage{textcomp}
\usepackage{lipsum}
\usepackage{amsmath}
\usepackage[numbers]{natbib}
\usepackage{url}
\usepackage[section]{placeins}
\usepackage{authblk}
\usepackage{float}
\usepackage{array}
\usepackage{tabularx}
\usepackage{caption}
\usepackage{adjustbox}
\newcolumntype{P}{>{\raggedright\arraybackslash}m{3.8cm}}
\newcolumntype{Q}{>{\raggedright\arraybackslash}m{4.2cm}}
\newcolumntype{R}{>{\raggedright\arraybackslash}m{9.25cm}}

\def\BibTeX{{\rm B\kern-.05em{\sc i\kern-.025em b}\kern-.08em
    T\kern-.1667em\lower.7ex\hbox{E}\kern-.125emX}}

\begin{document} 
\title{HIPAAChecker: The Comprehensive Solution for HIPAA Compliance in Android mHealth Apps \vspace{-4mm}
}

\author[1*]{Bilash Saha}
 \author[1]{Sharaban Tahora}
  \author[2]{Abdul Barek}
 \author[1]{Hossain Shahriar \vspace{-4mm}} 
\affil[1]{Department of Information Technology, Kennesaw State University, Georgia, United States}
\affil[2]{RightCodes Solutions, Dhaka, Bangladesh}


\affil[*]{{bsaha@students.kennesaw.edu} \vspace{-4mm}}

\vspace{-16mm}

\maketitle

\begin{abstract}
The proliferation of mobile health technology, or mHealth apps, has necessitated the paramount importance of safeguarding personal health records. These digital platforms afford individuals the ability to effortlessly monitor and manage their health-related issues, as well as store, share, and access their medical records and treatment information. As the utilization of mHealth apps becomes increasingly widespread, it is imperative to ensure that protected health information (PHI) is effectively and securely transmitted, received, created, and maintained in accordance with the regulations set forth by the Health Insurance Portability and Accountability Act (HIPAA). However, it is unfortunate to note that many mobile app developers, including those of mHealth apps, are not fully cognizant of the HIPAA security and privacy guidelines. This presents a unique opportunity for research to develop an analytical framework that can aid developers in maintaining a secure and HIPAA-compliant source code, while also raising awareness among consumers about the privacy and security of sensitive health information. The plan is to develop a framework which will serve as the foundation for developing an integrated development environment (IDE) plugin for mHealth app developers and a web-based interface for mHealth app consumers. This will help developers identify and address HIPAA compliance issues during the development process and provide consumers with a tool to evaluate the privacy and security of mHealth apps before downloading and using them. The goal is to encourage the development of secure and compliant mHealth apps that safeguard personal health information.
\end{abstract}


\section{Introduction}

\begin{table*}
\footnotesize
\centering
\setlength{\tabcolsep}{0.5em}
{\renewcommand{\arraystretch}{1.2}
 \begin{tabularx}{\textwidth}{|l|l|X|}
    \hline
        \textbf{Reference} & \textbf{Rule Name} & \textbf{ Technical Safeguards} \\ \hline
        164.312(a)(1)   & Authorization & Implement technological policies and procedures to restrict access to individuals or software programs that have been given access privileges for electronic information systems that maintain EPHI. \\ \hline
        164.312(a)(2)(i)  & Unique Id  & Assign a unique name or number to each patient in order to identify and monitor their identification. \\ \hline
        164.312(a)(2)(ii)  & Emergency EPHI Access &  Create and use processes for acquiring required digitally protected health information in an emergency. \\ \hline
        164.312(a)(2)(iii)   & Automatic Session timeout & Implement software procedures that end a session after a certain period of inactivity. \\ \hline
        164.312(a)(2)(iv)   & EPHI Encryption and Decryption & Implement a system for encrypting and decrypting EPHI. \\ \hline
        164.312(b)   & EPHI Audit Control & Implement methods for recording and examining activities in information systems that use or include EPHI.  \\ \hline
        164.312(c)(1)   & EPHI Data Integrity & Implement regulations and procedures to prevent unauthorized manipulation or destruction of EPHI. \\ \hline
        164.312(c)(2)   & EPHI Integrity Verification & Utilize technological tools to verify that electronically stored protected health information has not been tampered with or deleted without authorization. \\ \hline
        164.312(d)   & EPHI Authentication & Establish processes to confirm that the individual or organization requesting access to EPHI is who is being identified. \\ \hline
        164.312(e)(1)   & EPHI Transmission Security & Implement technological security measures to prevent unauthorized access to digitally protected health information that is being sent through a network of electronic communications. \\ \hline
        164.312(e)(2)(i)   & EPHI Transmission Integrity & Implement security measures to guarantee that electronically transmitted protected health information is not improperly altered up to disposal without being noticed. \\ \hline
        164.312(e)(2)(ii) & Appropriate EPHI Encryption  & Implement a mechanism to encrypt EPHI whenever deemed appropriate. \\ \hline
    \end{tabularx}%
}
\caption{HIPAA Technical Safeguards with Corresponding Rule Names}
\label{tab:1}
\vspace{-4mm}
\end{table*}

Due to the fact that mHealth systems acquire, process, store, and transport sensitive user data as well as individual health records, the situation with regard to security vulnerabilities is particularly crucial. In order to guarantee the confidentiality, integrity, and accessibility of electronic health information that is kept or transferred electronically, the HIPAA \cite{HIPAA} security rule, which went into effect in April 2005, imposes administrative, physical, and technical measures \cite{ref_1}, \cite{ref_2}. To safeguard patient and healthcare professional data, mHealth apps must be secured. According to a recent study in this area, security threats for mobile health applications might be divided into three categories: high (apps for monitoring, diagnosis, and care), medium (calculators, localizers, and alarms), and low (informative and educational apps) \cite{ref_3}. The American Health Information Management Association (AHIMA) provided advice \cite{ref_19} on how to deal with mobile health data breaches, including reviewing privacy settings on both apps and mobile devices, looking for certification signs, using password and encryption, and refraining from texting others with private or sensitive health information. To lessen security and privacy issues, the majority of vulnerabilities in the mobile health app should be addressed and resolved. Before permitting the usage of the mHealth applications, such initiatives need help to evaluate the source code and test them in accordance with the HIPAA's subsequent security and privacy criteria. According to\cite{ref_4}, the privacy and security of personal health records are key issues. The absence of standardized mHealth applications and security concerns are a major impediment to their broad deployment. A comparative study [5] \cite{ref_5} of the top 20 mHealth applications found that while just two apps needed user authentication before logging in, 65\% of the apps asked users to submit personal information such as name, address, email, and DOB. Data breaches and the privacy of users' personal information are seriously threatened by the 50\% of applications that store data on the cloud. Additionally, more than 65\% of applications shared user data with advertising or third parties without getting user permission, which is against the law. Only 20\% of applications provided users with information regarding data privacy and security measures. Authors provided a static security analysis method using the free and open-source FindSecurityBugs\cite{findbugs} IDE plugin for Android Studio. They showed how integrating the plugin helps developers to safeguard mobile applications and lessen security threats as they are being implemented. According to a thorough review of the literature and internet searches done as of March 2023,  \cite{DexGuard}, \cite{TrustKit}, \cite{RADIO} \cite{ref_15}, there aren't any tools or frameworks that verify the security of mHealth apps using the HIPAA security standards for EPHI. With the use of supplementary code analysis tools like FindBugs  \cite{findbugs}, IntelliJ IDE  \cite{intellij}, and Eclipse IDE, developers may maintain and tidy up their code. These tools are designed to find possible flaws like inconsistencies, aid in improving the structure of the code, conform the source code to standards, and offer rapid fixes. Their primary responsibility is not to check security risks based on HIPAA technical security criteria \cite{HIPAA}. A complete list of Android app analysis tools is provided in a study  \cite{ref_12}, however none of them concentrate on mHealth app security and privacy analysis in accordance with HIPAA technical security and privacy standards. Recent examples of mobile security analysis tools that do not concentrate on HIPAA violations are DexGuard \cite{DexGuard} and TrustKit \cite{TrustKit}. 

The aim of this research is to promote the development of secure and compliant mHealth apps that protect the confidentiality of personal health information. So the main objectives are as follows: 

\begin{itemize}
    \item Develop a source code analysis framework to evaluate HIPAA Technical Safeguards for determining the compliance of mHealth applications.
    \item Incorporate API-level checking in accordance with secure data communication between third-party mHealth apps and electronic health record systems. 
    \item Implementation of meta-analysis to identify potential risks and safety features, and in detecting HIPAA violations.
    \item Develop an IDE plugin tool that provides developers with analysis and feedback on their codebase to address potential security and privacy issues early in the development process.
\end{itemize}

\begin{figure}[H]
\centerline{\includegraphics[width=\columnwidth]{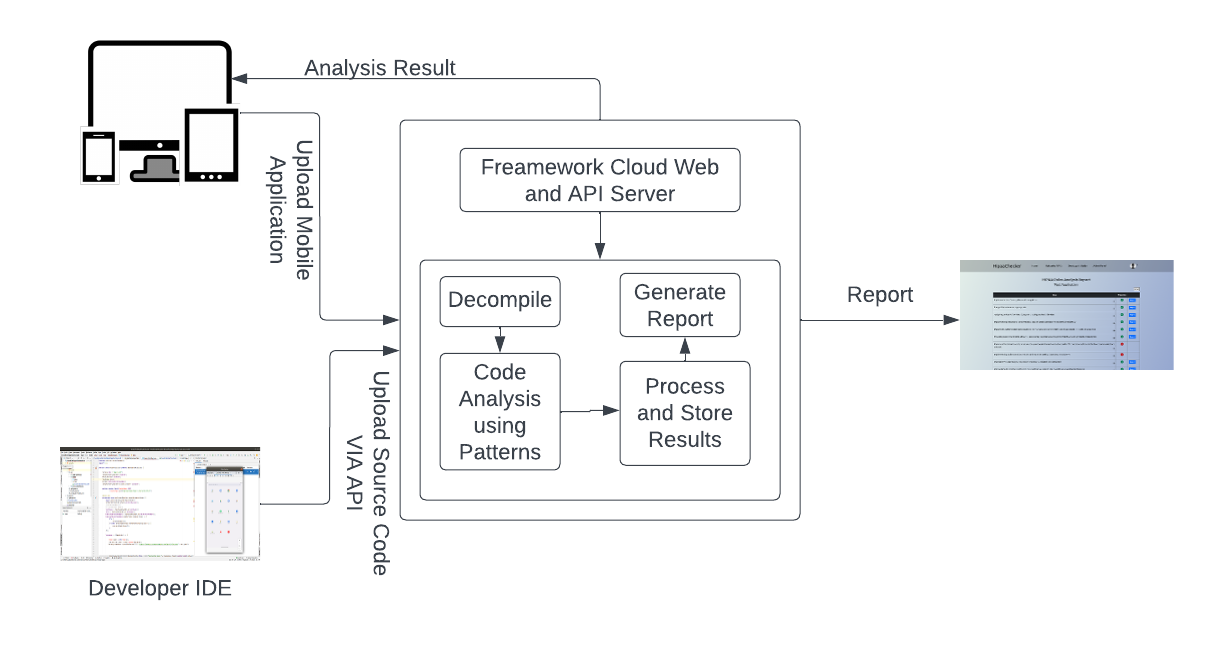}}
\caption{HIPAAChecker Framework Architecture}
\label{fig:architechture}
\end{figure}

\section{Research Methods}

Administrative, Physical, and Technical security requirements are outlined in the Health Insurance Portability and Accountability Act (HIPAA). Administrative safeguards are rules and guidelines that control the choice, creation, application, and upkeep of security measures. Physical safeguards are techniques, guidelines, and practices created to protect equipment from environmental and natural threats as well as unlawful access. Technical safeguards, on the other hand, refer to the policy- and technology-related procedures that defend against unauthorized access to electronically protected health information (EPHI) \cite{ref_20}, \cite{ref_21}, \cite{staff2022}. Proposed source code analysis approaches for mHealth apps specifically address problems with Technical Safeguards (Table \ref{tab:1})  \cite{safeguard}, \cite{staff2022}. It will be possible to meet other administrative and operational safeguards, such as providing tools and applications to review and monitor administrative security features, and prevent negative incidents, such as non-compliance with physical safeguards, by ensuring technical safeguard compliance. For instance, obtaining encrypted PHI data from a lost or stolen cell phone's mHealth app would be quite challenging.

\section{Hipaachecker Framework}

\begin{figure*}[!htbp]
\centerline{\includegraphics[width=1.0\textwidth]{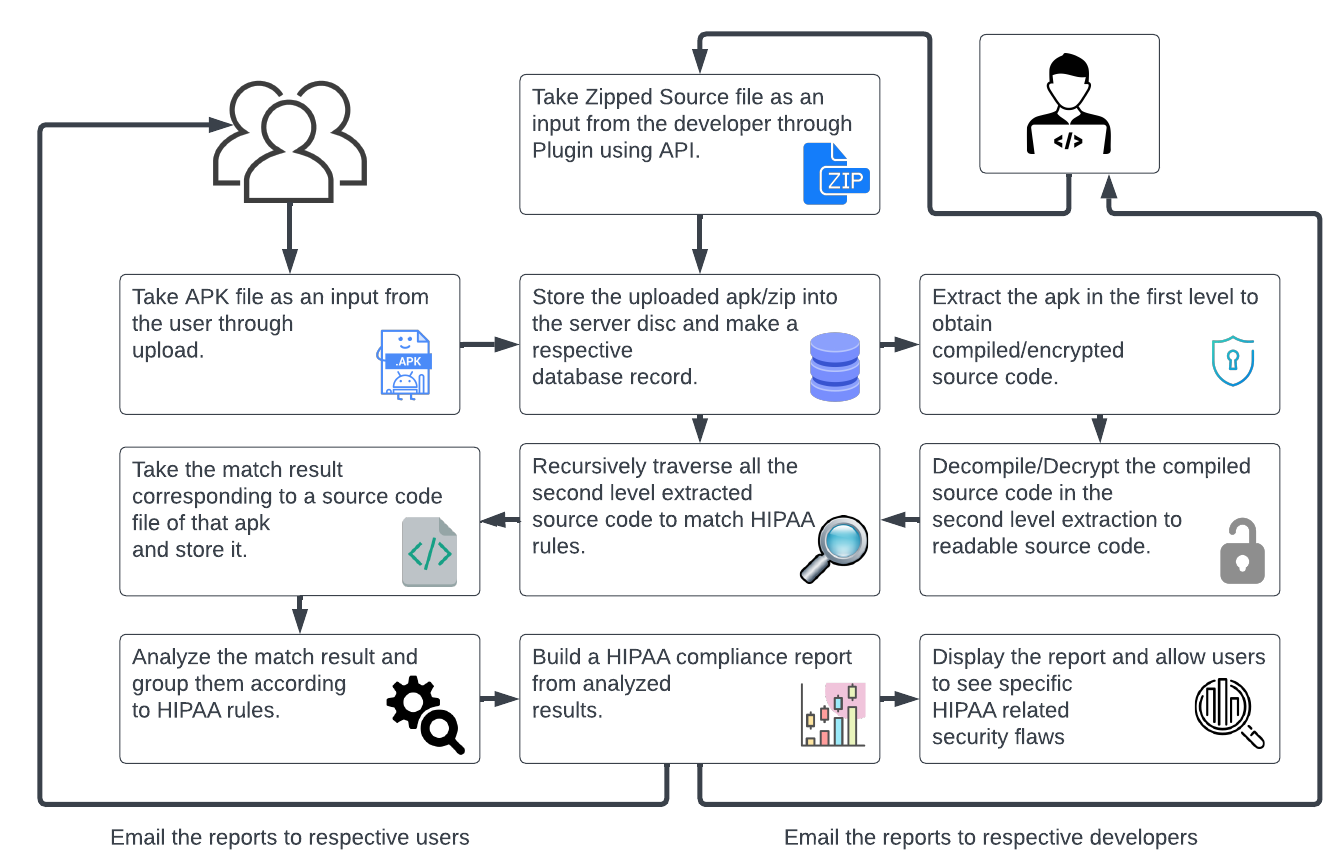}}
\caption{HIPAAChecker Framework's Meta Analysis Flow }
\label{fig:Flow}
\end{figure*}

\begin{figure}[!htbp]
\centerline{\includegraphics[width=\columnwidth]{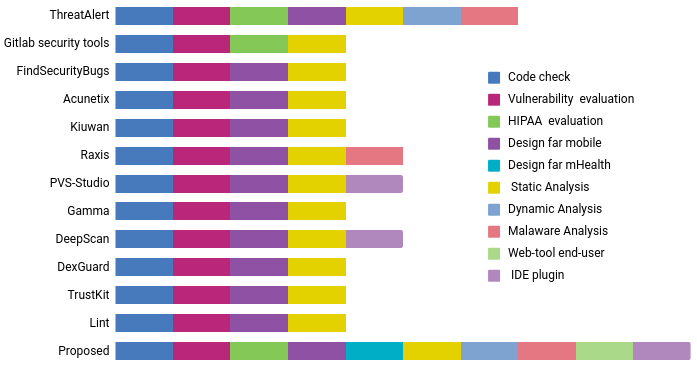}}
\caption{ Comparison of the features of the proposed framework with those of similar products on the market }
\label{fig:Comparison}
\end{figure}

In this study, a methodology for examining Android mHealth apps that are intended to accept patient data as input and follow HIPAA technical security criteria for data storage and transmission is proposed. The goal of this framework is to automatically evaluate application source code and discover security and privacy patterns prevalent in mHealth apps, in contrast to certain existing analysis tools that concentrate on Java-specific security checks. The framework's overall architecture is depicted in Figure \ref{fig:architechture}, and its properties are contrasted with those of other tools already in use in Figure \ref{fig:Comparison}. Before publishing apps through Android Studio to the market, the plugin tool enables mHealth developers to find and fix flaws that can affect HIPAA technical security and privacy requirements. Additionally, by submitting APK files, users of common mHealth applications can use the tool to look for security flaws. The meta analysis flow in relation to regular web users and developers is depicted in Figure \ref{fig:Flow}.

The framework uses JADX \cite{jadx} to decode manifest and other resources and decompile the dex file into Java classes because Android applications are often developed in Java, compiled into dex format files, and executed in instances of the Android virtual machine. A source code analyzer employing patterns is built into the framework cloud server machine. It will look for certain HIPAA-related code-level vulnerabilities and process the results to produce reports for the targeted users. The report will detail the precise lines of source code that need to be refactored to make the application HIPAA compliant.  

Overall, the proposed framework will give mHealth app developers a methodical way to check that their applications meet HIPAA's technical security criteria, thereby improving the privacy and security of patient data.

\subsection{Framework Development:}
\begin{figure}[!htbp]
\centerline{\includegraphics[width=\columnwidth]{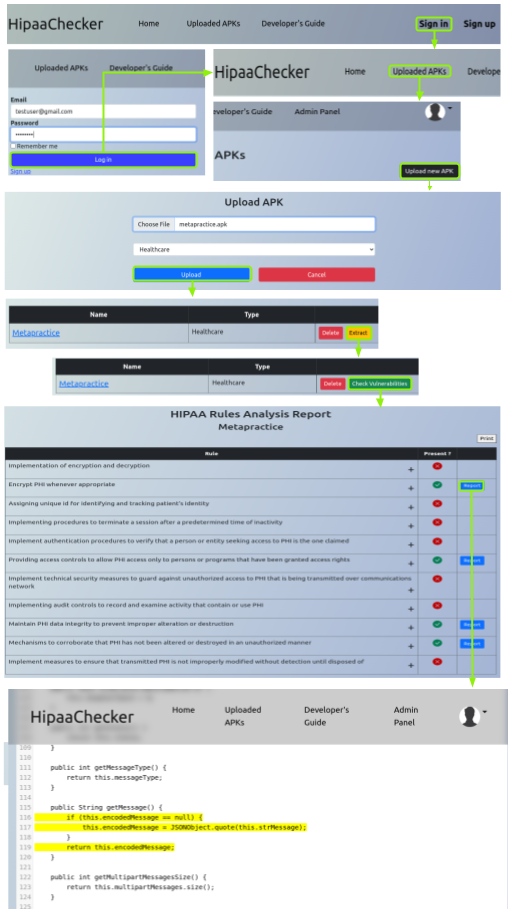}}
\caption{HIPAAChecker Web Application Report Generation via uploading APK file }
\label{fig:WebFlow}
\end{figure}

As per Figure \ref{fig:Flow}, We have developed a web-based platform for conducting experiments on HIPAA compliance. The web application of the framework allows users to verify HIPAA compliance via APK file URLs and generate reports. The platform features two layers of access control: Layer 1 involves username and password authentication, and Layer 2 involves user account and authentication security using two-factor authentication (2FA).
As shown in the Figure \ref{fig:WebFlow}, A user may sign up for the online application, and when their email has been verified, they can log in. Any APK file may thereafter be uploaded by the user. The system will store the submitted APK and create a corresponding database record after a successful upload. There will now be a button that says "Extract," and when users click it, the program will extract the APK in two steps: first, to get the compiled source code, and second, to get the readable source code. A button with the caption "Check Vulnerabilities" will show up upon successful extraction. The system will recursively search through the retrieved source code after users click this button to see whether any of the patterns shown in Table \ref{tab:patterns} are present. If a match has been found, the match findings will be stored in the database for further analysis. After the traversal is finished, all of the match results will be combined, and the user will be shown a report with a legible high label. There will be a green check mark in this report indicating that the uploaded application complies with the relevant HIPAA requirement. If the requirement is not satisfied in the submitted application, they will see a cross. The user may then click on a particular rule's name to learn more about its sub rules match. The user can click a specific "Report" button to view a code-level report that includes the line numbers of the relevant lines of code and a matching section of code that is connected to the relevant HIPAA regulation. They can click on the specific link of that page and see the full source code file and specific matched line highlighted. 

\begin{table*}[htbp!]
\footnotesize
\centering
\setlength{\tabcolsep}{0.5em}
{\renewcommand{\arraystretch}{1.2}
 \begin{tabularx}{\textwidth}{P | Q | R}
\hline
\rowcolor[HTML]{D0EEF7} 
\multicolumn{1}{|c|}{\cellcolor[HTML]{D0EEF7}\textbf{Rule ID}} &
  \multicolumn{1}{c|}{\cellcolor[HTML]{D0EEF7}\textbf{Sub Rule ID}} &
  \multicolumn{1}{c|}{\cellcolor[HTML]{D0EEF7}\textbf{Detection Code Patterns}} \\ \hline
 &
  EN-DE &
  import java.util Base64 \\ \cline{2-3} 
 &
  AES &
  import org.springframework.security.crypto, import java.security.Security;, Cipher.getInstance("AES/ECB/, Cipher.getInstance("AES"),Cipher.getInstance(AES\_MODE , new SecretKeySpec(keyBytes, "AES", Cipher.getInstance("AES/CBC/ \\ \cline{2-3}
 &
  DES &
  Cipher.getInstance(.*DES, Cipher.getInstance(.*des \\ \cline{2-3} 
 &
  RSA &
  Cipher.getInstance("RSA \\ \cline{2-3} 
 &
  BLOWFISH &
  .getInstance(.*BLOWFISH \\ \cline{2-3} 
 &
  RC &
  .getInstance(.*RC2, .getInstance(.*rc4, .getInstance(.*RC4, .getInstance(.*rc2 \\ \cline{2-3} 
 &
  Message Digest &
  MessageDigest, import java.security.MessageDigest;, .getInstance(.*MD5, .getInstance(.*md5, DigestUtils.md5(, import org.apache.commons.codec.digest.DigestUtils; \\ \cline{2-3} 
 &
  SHA &
  .getInstance(.*SHA-1, .getInstance(.*SHA1, DigestUtils.sha( \\ \cline{2-3} 
 &
  ECB &
  - Cipher.getInstance(\textbackslash{}s*"\textbackslash{}s*AES\textbackslash{}/ECB \\ \cline{2-3} 
\multirow{-10}{*}{EPHI\_encryption\_decryption} &
  HMAC &
  - import org.apache.commons.codec.digest.HmacAlgorithms;, import org.apache.commons.codec.digest.HmacUtils; \\ \hline
\rowcolor[HTML]{EFEFEF} 
EPHI\_Transmission\_integrity &
  TRANS-NET &
  javax.net.ssl.TrustManager, TrustManagerFactory.getInstance( \\ \hline
 &
  DE &
  android.util.Base64, .decodeToString, .decode \\ \cline{2-3} 
 &
  EN &
  android.util.Base64, .encodeToString, .encode \\ \cline{2-3} 
 &
  ENCRYPT &
  io.realm.Realm, .encryptionKey( \\ \cline{2-3} 
\multirow{-4}{*}{Appropriate\_ EPHI\_Encryption} &
  Chiper &
  net.sqlcipher., AS encrypted KEY \\ \hline
\rowcolor[HTML]{EFEFEF} 
\cellcolor[HTML]{EFEFEF} &
  Authorization Control &
  AuthorizationException \\ \cline{2-3} 
\rowcolor[HTML]{EFEFEF} 
\multirow{-2}{*}{\cellcolor[HTML]{EFEFEF}Authorization} &
  Access Control &
  IllegalAccessException \\ \hline
 &
  API &
  addRequestProperty(\textbackslash{}"Authorization \\ \cline{2-3} 
 &
  PKIX &
  PKIXRevocationChecker \\ \cline{2-3} 
\multirow{-3}{*}{EPHI\_Transmission\_Security} &
  TRANS-Data &
  HttpsURLConnection new \\ \hline
\rowcolor[HTML]{EFEFEF} 
Unique\_Id &
  PK &
  PRIMARY KEY \\ \hline
 &
  FireBaseAuth &
  FirebaseUser, sendFirebasePropertyRegisteredUser, FirebaseUserPropertiesSender, com.google.firebase\textbackslash{}:firebase-auth, FirebaseAuth \\ \cline{2-3} 
\multirow{-2}{*}{EPHI\_authentication} &
  aAuth &
  android.accounts.AccountManager, AccountManager.get(, .currentUser \\ \hline
\rowcolor[HTML]{EFEFEF} 
Automatic\_Session\_Timeout &
  Inactivity &
  public void onUserInteraction(), .reset(), .clear(), .commit() \\ \hline
EPHI\_Audit\_Control &
  Audit &
  AppOpsManager.OnOpNotedCallback \\ \hline
\rowcolor[HTML]{EFEFEF} 
\cellcolor[HTML]{EFEFEF} &
  authorization\_exception\_on\_destroy &
  AuthorizationException \\ \cline{2-3} 
\rowcolor[HTML]{EFEFEF} 
\multirow{-2}{*}{\cellcolor[HTML]{EFEFEF}EPHI\_integrity\_verification} &
  illegal\_destruction\_restriction &
  IllegalAccessException \\ \hline
 &
  authorization\_exception &
  AuthorizationException \\ \cline{2-3} 
 &
  illegal\_access &
  IllegalAccessException \\ \cline{2-3} 
 &
  user\_authentication\_oauth &
  android.accounts.AccountManager, AccountManager.get( \\ \cline{2-3} 
\multirow{-4}{*}{EPHI\_data\_integrity} &
  user\_authentication\_firebase &
  FirebaseUser, sendFirebasePropertyRegisteredUser, FirebaseUserPropertiesSender, com.google.firebase\textbackslash{}:firebase-auth, FirebaseAuth \\ \hline
\end{tabularx}%
}
\caption{Detection Code Patters corresponding to HIPAA rule ID and Sub Rule ID}
\label{tab:patterns}
\end{table*}

\begin{figure}[!htbp] 
\centerline{\includegraphics[width=\columnwidth]{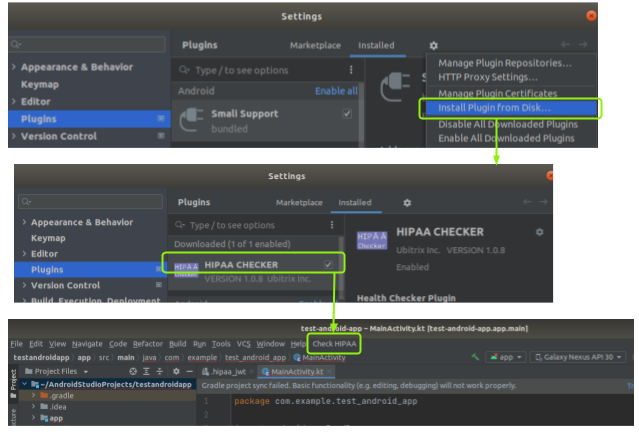}}
\caption{HIPAAChecker Web Application Report Generation via uploading APK file }
\label{fig:5}
\end{figure}

 Furthermore, we developed an IDE plugin that connects the analytical API to Integrated Development Environments (IDEs), such Android Studio. The "Check HIPAA" button that this plugin provides to the IDE enables developers to easily check if their code complies with HIPAA regulations. As seen in Figure \ref{fig:5}, when the button is pressed, the source code is compressed and uploaded to a centralized server. Following the code's being processed, as shown in Figure \ref{fig:Flow}, the server emails the developer a full report on the code's HIPAA compliance status. This way, the resource-intensive processing will be offloaded to the server, which relieves the burden on the developers' own machines. This facilitates developers' ability to resolve HIPAA-related security issues throughout the development process in their preferred development environment, enhancing the overall quality and compliance of the code. Our HIPAAChecker report provides developers with a granular level detailed analysis of their app's source code, highlighting specific line numbers where HIPAA compliance may be lacking. This greatly minimizes the work required of developers to ensure that their healthcare applications adhere to HIPAA regulations, increasing the effectiveness and efficiency of development.

\subsection{Testing and Evaluation }
We have manually downloaded a collection of 285 mHealth apps from the Medical and Health \& Fitness categories of the Google Play Store and Github to evaluate the potential threat of HIPAA violation posed by current health apps. The selection of the apps was based on the prioritization of features and functionalities that involve the storage, processing, management, with consideration given to covered entities and business associates of the selected apps from different geographical locations. Additionally, we took into account the privacy policy, terms of service, collection techniques, and transfer of EPHI.

\begin{figure}[!htbp]
\centerline{\includegraphics[width=\columnwidth]{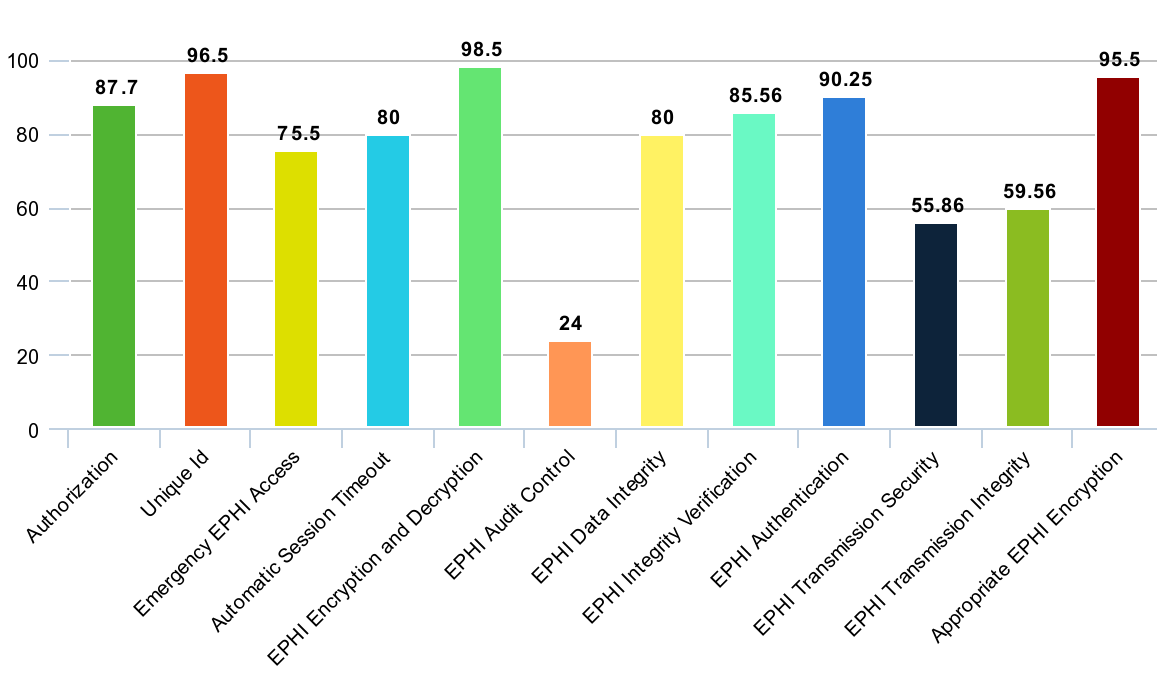}}
\caption{Percentage of apps that met the HIPAA technical safeguards }
\label{fig:6}
\end{figure}

\begin{figure}[!htbp]
\centerline{\includegraphics[width=\columnwidth]{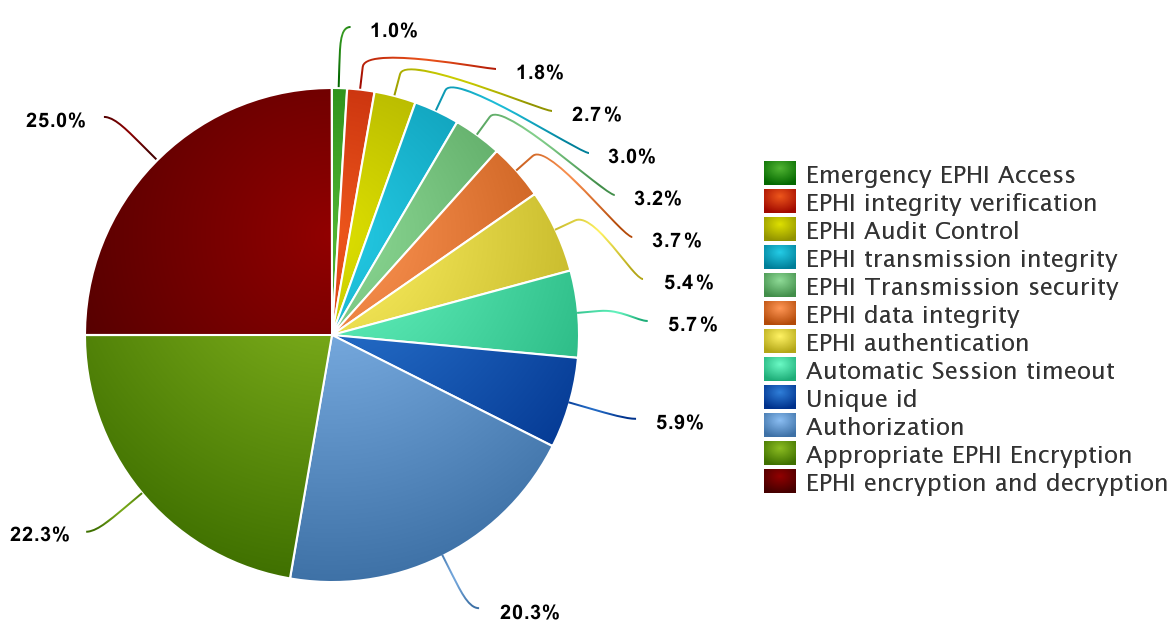}}
\caption{Percentage of code segments detected in all downloaded apps that comply to HIPAA regulations}
\label{fig:7}
\end{figure}

\begin{figure}[!htbp]
\centerline{\includegraphics[width=\columnwidth]{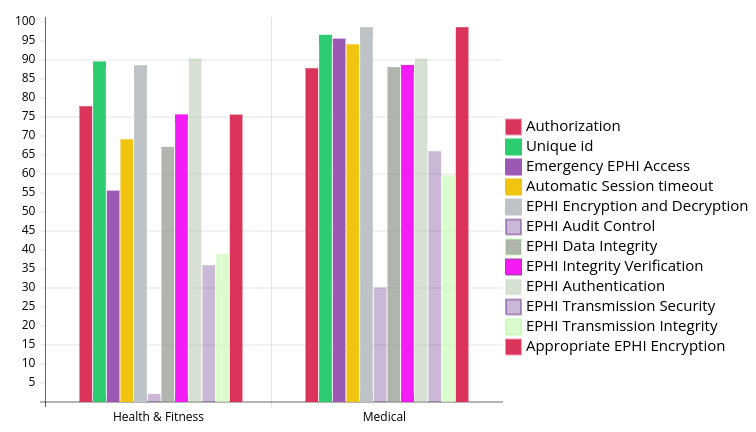}}
\caption{Percentage of apps that met the HIPAA technical safeguards with respect
to apps categories }
\label{fig:8}
\end{figure}

In addition, we have taken into account both top-rated downloaded apps (10M+) and low-rating apps (100+ downloads) during the selection process. We have tested the Google Play Store applications through our developed web application, and the Github open-source repositories have been tested via the developed IDE plugin. As illustrated in Figure \ref{fig:6}, our investigation demonstrated that a sizable portion of the downloaded apps lacked appropriate audit control or transmission security measures. Additionally, as shown in Figure \ref{fig:8}, we discovered that medical applications were more HIPAA compliant than health and fitness applications.
However, there were some encouraging results as well. As shown in Figure \ref{fig:7}, the majority of the apps that we examined implemented adequate authorization, unique user identification, and the majority of the apps used EPHI encryption/decryption mechanisms. The graphic representations of our findings are shown in Figures \ref{fig:6}, \ref{fig:7}, and \ref{fig:8}.

\section{Recommendations}
Our analysis report shows that the main hazards of these applications are lack of audit control, unsecured information sharing with third-party APIs, and unauthorized access to critical resources.
Table \ref{tab:recommendations} contains a collection of recommendations for both app developers and customers.

\begin{table}[!h]
\centering
\footnotesize
\begin{adjustbox}{width=0.50\textwidth}
\begin{tabular}{|l|}
\hline
\rowcolor[HTML]{329A9D} 
\multicolumn{1}{|c|}{\cellcolor[HTML]{329A9D}\textbf{Application Users}} \\ \hline
\rowcolor[HTML]{D5F4F5} 
\begin{tabular}[c]{@{}l@{}}- Check the application's review and privacy policy. \\   - Take advantage of tools to confirm HIPAA compliance before \\      disclosing personal health information.\\   - Examine and criticize apps, To inform others and assist researchers \\      and developers in effectively rebuilding the app.\end{tabular} \\ \hline
\rowcolor[HTML]{D3D352} 
\multicolumn{1}{|c|}{\cellcolor[HTML]{D3D352}\textbf{Application Developers}}     \\ \hline
\rowcolor[HTML]{F5F5C3} 
\begin{tabular}[c]{@{}l@{}}- Implement audit controls to enable thorough investigation of every incident.\\  - When integrating external APIs, be sure to use SSL.\\  - Use appropriate access control methods to ensure that only authorized \\    individuals can access sensitive EPHI.\\  - Use cutting-edge encryption and decryption mechanisms\end{tabular} \\ \hline
\end{tabular}%
\end{adjustbox}
\caption{Recommendations for Application users and developers}
\label{tab:recommendations}
\end{table}

\vspace{-1mm}

\section{Conclusion}
The use of mHealth applications is widespread, yet many of them have security and privacy flaws and don't adhere to HIPAA Technical Safeguard requirements. Our developed HIPAAChecker can solve this issue by spotting the absence of technical security measures in both released and under-development applications. By guaranteeing that applications are in compliance with HIPAA regulations, this framework seeks to increase the trust of application users. Using our tool, developers can find and fix any security or privacy flaws in their applications. By incorporating HIPAA safety measures, the healthcare and fitness industry can improve the security of sensitive EPHI and build trust between patients and healthcare providers.

\section*{Acknowledgment}
\footnotesize
This work is partially supported by National Institute of Health (NIH) under STTR award \#R41GM146313, and National Science Foundation (NSF) under awards \#2100115 and \#2209638.

\renewcommand{\bibfont}{\footnotesize}

\bibliographystyle{IEEEtranN}
\bibliography{biblography.bib}
\end{document}